# The Insecure Future of the World Economic Growth

Ron W Nielsen[1]

Environmental Futures Research Institute, Gold Coast Campus, Griffith University, Qld, 4222, Australia

November, 2015

Growth rate of the world Growth Domestic Product (GDP) is analysed to determine possible pathways of the future economic growth. The analysis is based on using the latest data of the World Bank and it reveals that the growth rate between 1960 and 2014 was following a trajectory approaching asymptotically a constant value. The most likely prediction is that the world economic growth will continue to increase exponentially and that it will become unsustainable possibly even during the current century. A more optimistic but less realistic prediction is based on the assumption that the growth rate will start to decrease linearly. In this case, the world economic growth is predicted to reach a maximum, if the growth rate is going to decrease linearly with time, or to follow a logistic trajectory, if the growth rate is going to decrease linearly with the size of the world GDP.

**Introduction**

Earlier studies (Nielsen, 2014, 2015a) indicated that the world economic growth increases too fast and that it might become unsustainable. We now investigate this issue further by using the empirically-determined growth rate, which offer a better way of studying the past and future trends. Our analysis is supported by the World Bank data (World Bank, 2015) of the Gross Domestic Product (GDP). Unlike the data of Maddison (2010), which are terminated at 2008, the World Bank data extend to 2014 and are continually updated. Maddison's data are useful for studying the historical economic growth because they extend down to AD 1 but the World Bank data are more useful for studying the modern economic growth.

**Mathematical methods**

Mathematical methods of investigation of growth trends have been described earlier (Nielsen, 2015b). Briefly, they are based on the mathematical analysis of the empirically-determined growth rates, $R$, which can be represented either as a function of time

$$R = \frac{1}{S}\frac{dS}{dt} = f(t) \qquad (1)$$

or as a function of the size $S$ of the growing entity

---

[1]AKA Jan Nurzynski, r.nielsen@griffith.edu.au; ronwnielsen@gmail.com; http://home.iprimus.com.au/nielsens/ronnielsen.html

Suggested citation: Nielsen, R. W. (2015). The Insecure Future of the World Economic Growth. http://arxiv.org/ftp/arxiv/papers/1510/1510.07928.pdf

For a broader discussion of the trends shaping the future of our planet see Nielsen (2006), the book endorsed by Nobel Laureate, Prof. Dr Paul Crutzen, translated into Chinese and recently reviewed by Chen (2013).

$$R = \frac{1}{S}\frac{dS}{dt} = f(S). \qquad (2)$$

In the first case, the solution of the differential equation (1) can be expressed using a general formula.

$$S(t) = \exp\left[\int f(t)dt\right]. \qquad (3)$$

In the second case, there is no such general description and different methods have to be used to solve the eqn (2). Furthermore, the eqn (2) has to be often solved numerically.

Empirical growth rate can be determined using the gradient calculated directly from data or by polynomial interpolation. It is generally essential to use interpolated gradient because in order to reproduce the past growth and to predict the future we have to understand the growth trajectory of the growth rate.

We shall now present the analysis of the world economic growth based on two assumptions: (1) that the world economic growth will continue to increase along the trajectory determined by the stable growth-rate trajectory of the past 54 years (between 1960 and 2014) and (2) that the economic growth will be now diverted to a new trajectory determined by a new, linearly-decreasing, growth-rate trajectory. The first assumption is more realistic because it is based on using the whole range of growth rate data. The second assumption is less realistic but its outcome more desirable because linearly-decreasing growth rate leads either to a maximum or to a levelling off of the economic growth.

**Realistic predictions**

Growth rates of the world GDP calculated directly from data (World Bank, 2015) and by using interpolated gradient are shown in Figure 1. As expected, the growth rate calculated directly from data, *R* (*Direct*), shows strong fluctuations, which are the reflection of the fluctuations in the gradient. They disappear when the interpolated gradient is used and they reveal a clear and well-defined trend, which can be analysed mathematically.

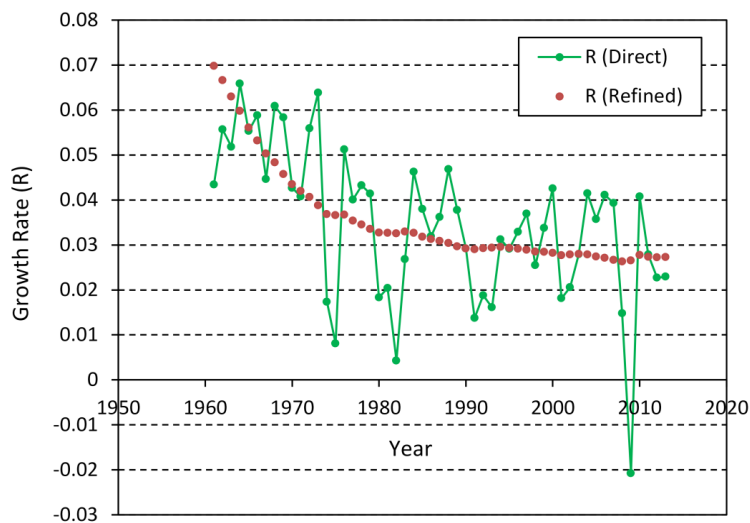

Figure 1. Empirical growth rate of the world GDP calculated directly from data (World Bank, 2015) [*R* (*Direct*)] and by using interpolated gradient [*R* (*Refined*)].

Empirical growth rate [R (*Refined*)] follows the reciprocal values of a familiar saturation distribution (Nielsen, 2011), which approaches asymptotically a constant value:

$$R = (a - be^{-rt})^{-1}. \tag{4}$$

It is always useful to try to reduce mathematical analysis to a straight line because a curve fitting data can be then more convincingly identified. Let us replace the growth rate $R$ by the function $F$:

$$F \equiv a - \frac{1}{R}. \tag{5}$$

If the empirical growth rate follows the distribution given by the eqn (4) then the quantity $F$ representing the empirical growth rate $R$ should follow a decreasing exponential distribution:

$$F = be^{-rt}, \tag{6}$$

which means that $\ln F$ should follow a straight line

$$\ln F = \ln b - rt \tag{7}$$

Natural logarithm, $\ln F$, calculated using the empirical growth rate $R$ is shown in Figure 2. The empirical values of $\ln F$ follow closely the linear trend, confirming that the empirical values of the growth rate, $R$, follow the distribution described by the equation (4). However, significant deviations from the straight line towards the end of this trend give a slim hope that maybe in the future the empirical growth rate will be diverted to a slower trajectory. There was briefly such a trend when $\ln F$ values were below the fitted straight line but it was quickly compensated by new data, which are consistently above the straight line. Consequently, it is equally likely that the growth rate will continue along it asymptotic trajectory or that it might be even increasing, which would lead even faster to the global economic crisis.

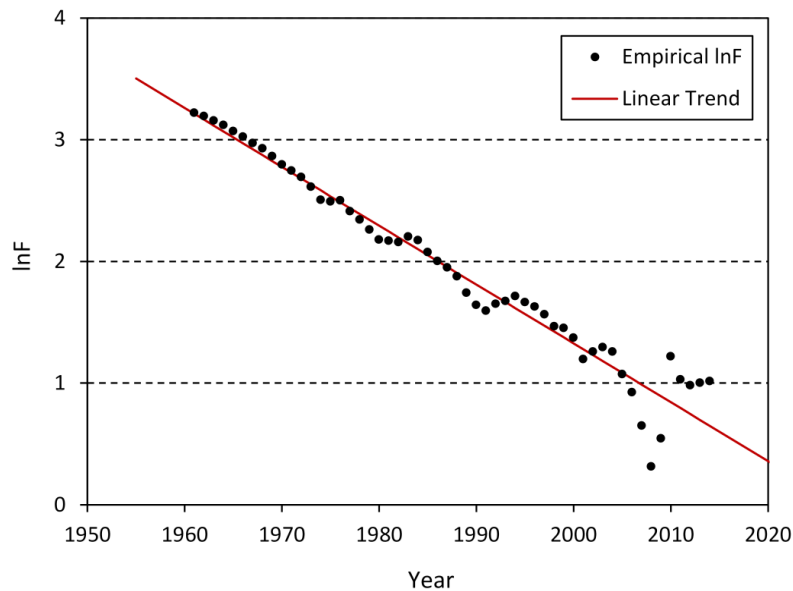

Figure 2. Empirical values of $\ln F$, calculated using the empirical growth rate, [R (*Refined*)], are compared with the best fit using linear function confirm that the growth rate, $R$, between 1960 and 2014 followed the trend described by the eqn (4).

The trajectory described by the eqn (4) is shown in Figure 3. It approaches asymptotically the limit

$$R_\infty = \frac{1}{a}. \qquad (8)$$

Constant growth rate describes exponential growth. The current growth rate is already close to its constant limit so even now it is approximately exponential. Such a growth is unsustainable.

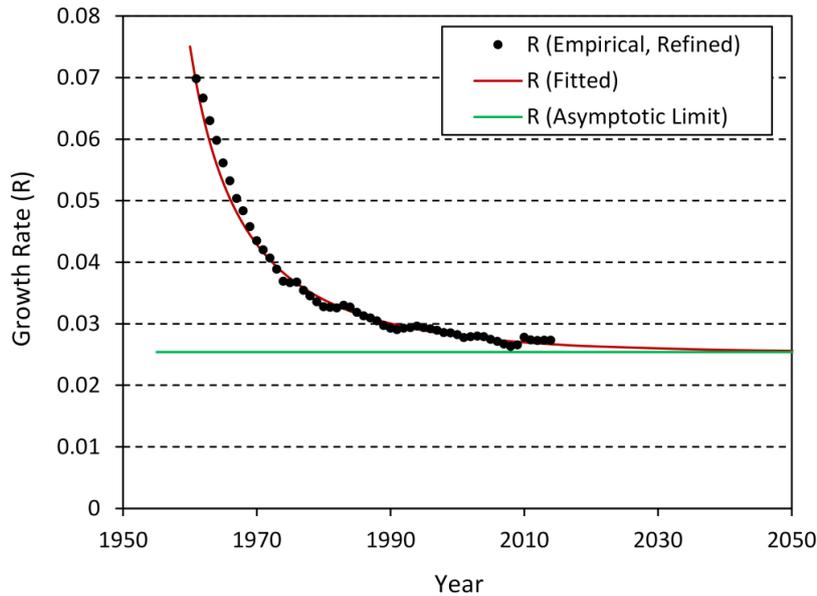

Figure 3. The empirical growth rate [$R$ (*Empirical, Refined*)] of the world GDP, expressed as a function of time, is compared with the distribution described by the eqn (4) incorporating the empirically-determined parameters *a*, *b* and *r*.

The parameters describing the fitted distribution [$R$ (*Fitted*)] are: $a = 3.940 \times 10^1$, $b = 3.787 \times 10^{42}$ and $r = 4.836 \times 10^{-2}$. The asymptotic limit of the growth rate is $2.538 \times 10^{-2}$ or approximately 2.5%. The growth rate in 2014 was 2.7%.

In order to find the mathematical description of the GDP trajectory and to explore its future trend we have to solve the following integration [see the eqns (1), (3) and (4)]:

$$\int f(t)dt = \int \frac{dt}{a - be^{-rt}}. \qquad (9)$$

Let us consider the general integral:

$$\int \frac{dx}{a + be^{rx}}. \qquad (10)$$

We can perform the integration by substitution

$$u = a + be^{rx} \qquad (11)$$

and by using the general integration formula we have found earlier (Nielsen, 2015b):

$$\int \frac{dx}{(a+bx)(c+ex)} = \frac{1}{\Delta} \ln \frac{a+bx}{c+ex}, \tag{12}$$

where

$$\Delta = cb - ae. \tag{13}$$

Consequently,

$$\int \frac{dx}{a+be^{rx}} = \frac{x}{a} - \frac{1}{ra} \ln(a+be^{rx}). \tag{14}$$

If we apply this formula to the integral (9) we shall find that,

$$\int \frac{dt}{a-be^{-rt}} = \frac{t}{a} + \frac{1}{ra} \ln(a-be^{-rt}). \tag{15}$$

Therefore,

$$S(t) = C \exp\left[\frac{t}{a} + \frac{1}{ra} \ln(a-be^{-rt})\right], \tag{16}$$

where $C$ is the constant of integration, which can be determined empirically by normalising the calculated curve to data.

We can see that

$$S(t \to \infty) \to C \exp\left(\frac{t}{a}\right). \tag{17}$$

Asymptotic constant $C$ is not the same as the constant $C$ in the eqn (16). The size of the GDP approaches asymptotically exponential distribution with the growth rate of $1/a$ and the doubling time $T_2 = a \ln(2)$.

The GDP distribution [*GDP (Best Fit)*] calculated using the eqn (16) and the empirically-determined parameters $a$, $b$ and $r$, is compared with the GDP data in Figures 4 and 5. This distribution is now merging with the exponential distribution, [*GDP (Asymptotic Exponential)*] characterised by the asymptotic limit, $1/a$, of the growth rate shown in Figure 3. Its doubling time is 27.3 years. If the growth is going to continue along the fitted trajectory, the predicted GDP in 2100 is $546 trillion 2005 US$ or 9.4 times higher than in 2014.

We shall now consider an alternative approach to fitting data and predicting future trend based on expressing the growth rate as a function of the *size* of the GDP rather than on time. In order to fit the data and to calculate future trend we now have to use the differential eqn (2) rather than the eqn (1).

The dependence of the empirical growth rate on the size of the GDP is shown in Figure 6 together with the fitted curve given by

$$R = \left(a - be^{-rS}\right)^{-1}. \tag{18}$$

The eqn (18) is a similar to the eqn (4) but now the dependence on time, $t$, is replaced by the dependence on the size, $S$, of the growing entity. In our analysis, $S$ represents the size of the GDP. For the large value of $S$, the growth rate approaches a constant value of $1/a$. The growth becomes exponential. The empirical growth rate in 2014 was close to its asymptotic limit. If the trend continues the world economic growth will continue along the ever-increasing exponential trajectory.

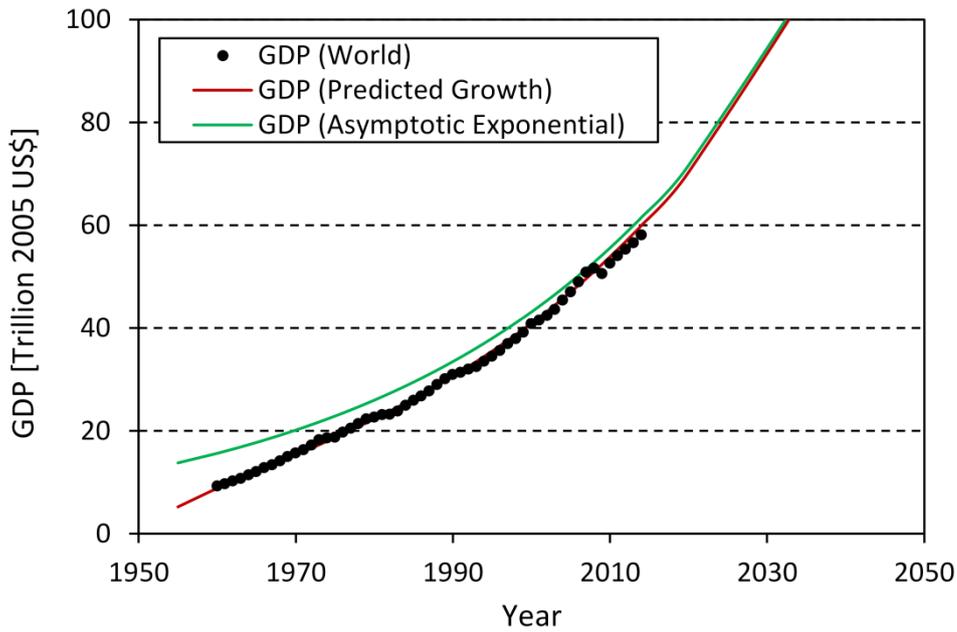

Figure 4. Predicted growth of the world GDP, [*GDP* (*Predicted Growth*)], calculated using the eqn (16) and the empirically-determined parameters *a*, *b* and *r* is compared with the GDP data (World Bank, 2015). Exponential distribution, [*GDP* (*Asymptotic Exponential*)], was calculated using the asymptotic value of $1/a = 2.538 \times 10^{-2}$.

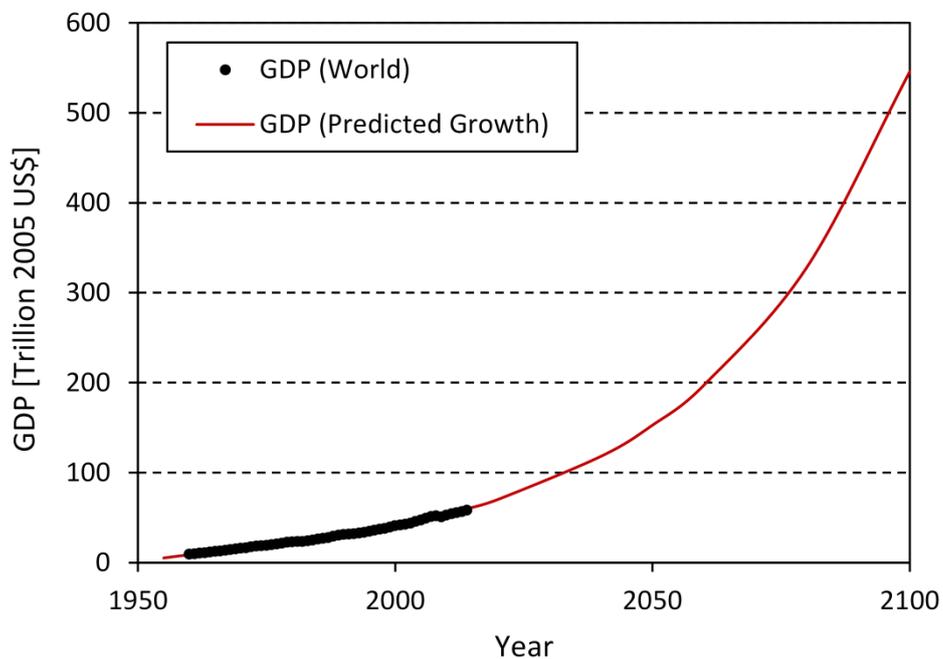

Figure 5. Predicted growth of the world GDP, [*GDP* (*Predicted Growth*)], is extended to 2100.

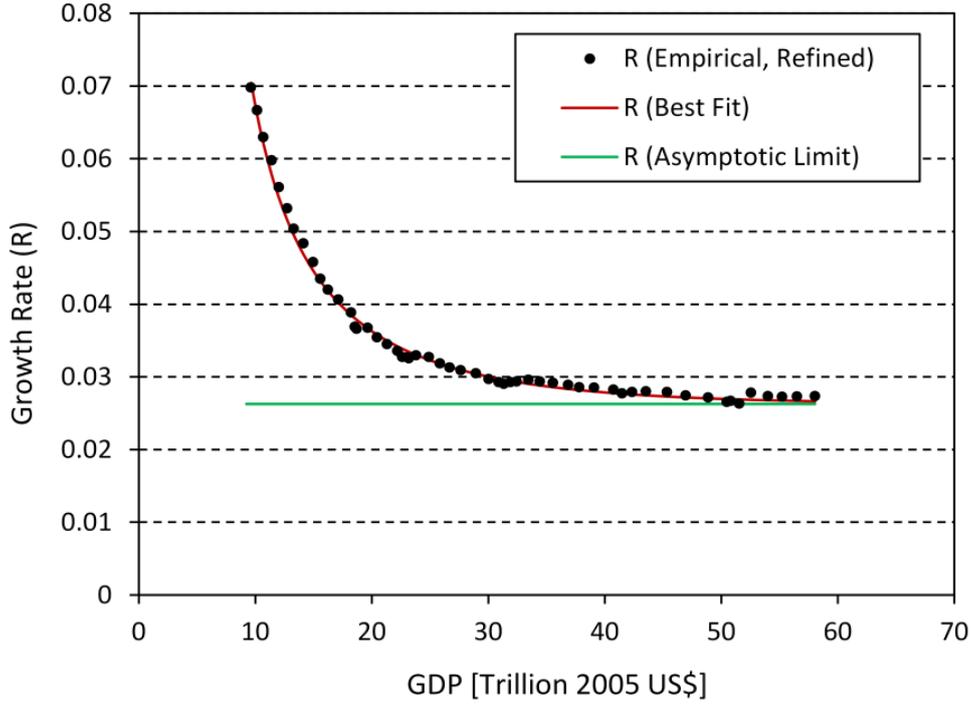

Figure 6. Empirical growth rate, [*R* (*Empirical, Refined*)], of the world GDP is plotted as a function of the *size* of the world GDP and compared with the distribution given by the eqn (18). The growth rate approaches asymptotically the limit given by the $1/a$ value.

To find the growth trajectory corresponding to the growth rate given by the eqn (18), we have to solve the following differential equation [see the eqn (2)]:

$$R = \frac{1}{S}\frac{dS}{dt} = \left(a - be^{-rS}\right)^{-1}, \qquad (19)$$

i.e. the equation

$$\frac{a - be^{-rS}}{S} dS = dt. \qquad (20)$$

Integration of the left-hand side of the eqn (20) gives

$$\int \frac{a - be^{-rS}}{S} dS = a \ln S - b \int \frac{e^{-rS}}{S} dS. \qquad (21)$$

The integral on the right-hand side can be found by using the series:

$$e^{qx} = 1 + qx + \frac{(qx)^2}{2!} + \frac{(qx)^3}{3!} + \frac{(qx)^4}{4!}\ldots \qquad (22)$$

Consequently,

$$\int \frac{a - be^{-rS}}{S} dS = (a-b)\ln S + b\left(rS - \frac{(rS)^2}{2\cdot 2!} + \frac{(rS)^3}{3\cdot 3!} - \frac{(rS)^4}{4\cdot 4!}\ldots\right). \qquad (23)$$

Solution to the differential eqn (20) is then given by

$$(a-b)\ln S + b\left(rS - \frac{(rS)^2}{2\cdot 2!} + \frac{(rS)^3}{3\cdot 3!} - \frac{(rS)^4}{4\cdot 4!}...\right) = t + C. \qquad (24)$$

This solution is an example of solutions, which cannot be used to unravel $S(t)$. In order to find $S(t)$ we have to solve the eqn (19) numerically using the parameters $a$, $b$ and $r$ determined by fitting the function given by the eqn (18) to the empirically-determined growth rate $R$. The parameters fitting the empirically-determined growth rate are: $a = 3.805 \times 10^1$, $b = 5.124 \times 10^1$ and $r = 7.927 \times 10^{-2}$. Again, this fit can be confirmed by using $\ln F$ defined by the eqn (5).

Numerical solution of the eqn (19) is displayed in Figure 7. The calculated distribution gives excellent fit to the GDP data and it also merges asymptotically with the exponential trajectory.

Distributions given by the eqns (4) and (18) appear to be complicated but in principle they are simple because they can be reduced to linear distributions represented by $\ln F$, where $F$ is defined by the eqn (5). They are also the simplest distributions fitting the growth rate data.

If it is possible to find a linear representation of data, then it is usually possible to identify their unique mathematical description. For instance, for a good quality data and for a sufficiently long range, linear representation of data obtained by calculating the logarithm of the studied quantity identifies exponential growth. Likewise, the linear representation of the reciprocal values of data identifies the first-order hyperbolic growth. Such linear representations of data help not only in the unique identification of growth trends but also in studying even small deviations from such trends because deviations from straight lines are easy to notice.

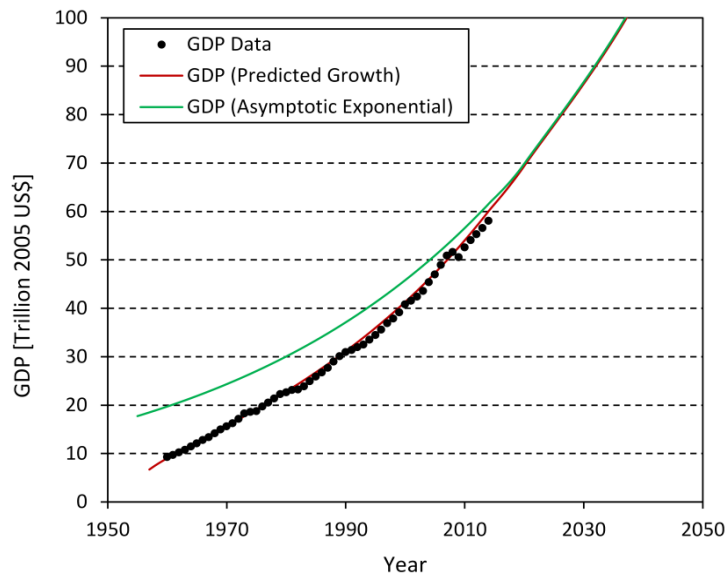

Figure 7. Predicted growth of the world GDP, [*GDP* (*Predicted Growth*)], represented by the numerical solution of the eqn (19) using the empirically-determined parameters $a$, $b$ and $r$ obtained by fitting the distribution defined by the eqn (18) to the *size*-dependent growth rate shown in Figure 6. The calculated distribution approaches asymptotically exponential growth, [*GDP* (*Asymptotic Exponential*)], characterise by the growth rate $1/a$.

**Preferred trajectories**

We have shown that the economic growth determined by the empirical growth rate between 1960 and 2014 is now approximately exponential. Such a growth is unsustainable over a sufficiently long time and at a certain stage it has to be terminated. For a sustainable economic growth, the growth rate should be now decreasing linearly because such a growth leads to a maximum, if the growth rate decreases linearly with *time*, or to a certain maximum level, if the growth rate decreases linearly with the *size* of the GDP (Nielsen, 2015b). We shall now explore this possibilities by assuming that the current asymptotically decreasing growth rate could be replaced by *linear approximations* determined by fitting growth rate data between 1980 and 2014 as shown in Figure 8.

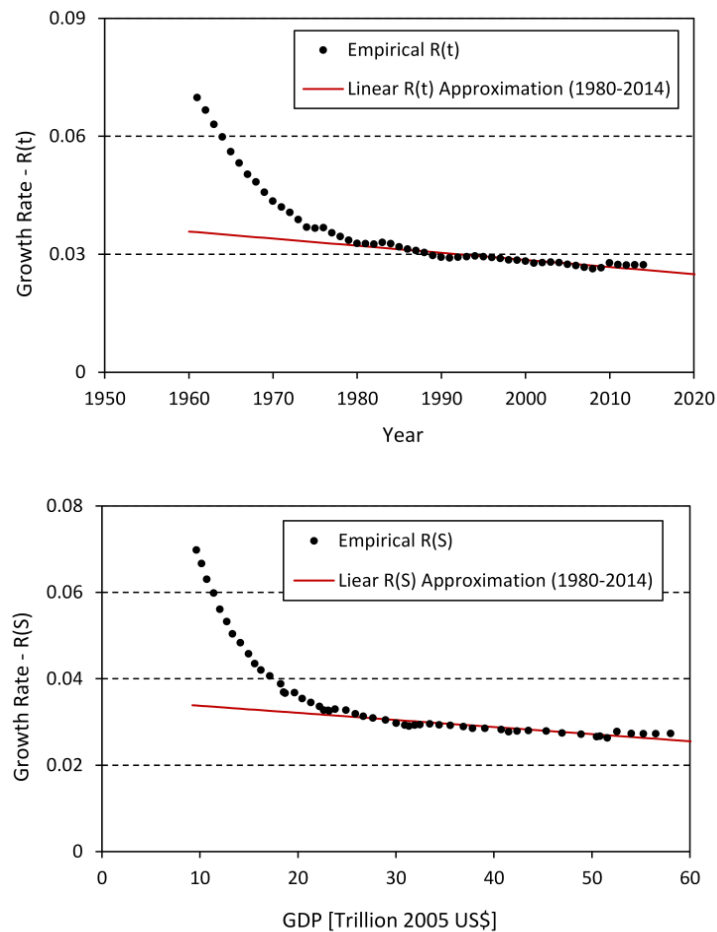

Figure 8. Growth rate of the world GDP is represented as a function of time, $R(t)$, or as the function of the size of the GDP, $R(S)$. The growth rate is approximated by fitting linear trends between 1980 and 2014.

For the growth rate represented by the linear trend depending on *time*,

$$R(t) = a + bt \qquad (25)$$

the empirically-determined parameters are $a = 3.895 \times 10^{-1}$ and $b = -1.805 \times 10^{-4}$.

For the growth rate represented by the linear trend depending on the *size S* of the world GDP,

$$R(S) = a + bS \qquad (26)$$

the empirically-determined parameters are $a = 3.539 \times 10^{-2}$ and $b = -1.641 \times 10^{-4}$.

As discussed earlier (Nielsen, 2015b), the growth rate given by the eqn (25) generates the trajectory

$$S(t) = C \exp\left[at + 0.5bt^2\right], \qquad (27)$$

which for $b < 0$ reaches a maximum at the time $t = a/|b|$.

The growth rate given by the eqn (26) generates a logistic-type of growth,

$$S(t) = \left[Ce^{-at} - \frac{b}{a}\right]^{-1}, \qquad (28)$$

which for $b < 0$ approaches asymptotically a limit of $S_\infty = a/|b|$.

These two, less-likely, growth trajectories described by the eqns (27) and (28) and by the empirically-determined parameters *a* and *b* are shown in Figure 9. They are compared with the more-likely growth trajectory determined using the full range of data for the growth rate between 1960 and 2014. The time-dependent linear approximation of the growth rate shown in the top section of Figure 8 generates the distribution leading to a maximum of $380 trillion 2005 US$ in 2158. The GDP-dependent linear approximation of the growth rate, shown in the lower section of Figure 8, generates the logistic-type of growth with the asymptotic limit of $216 trillion 2005 US$. A few examples of predicted values of the GDP and of the associated characteristic features of the projected trajectories are presented in Table 1. The corresponding empirically-determined parameters are listed in Table 2.

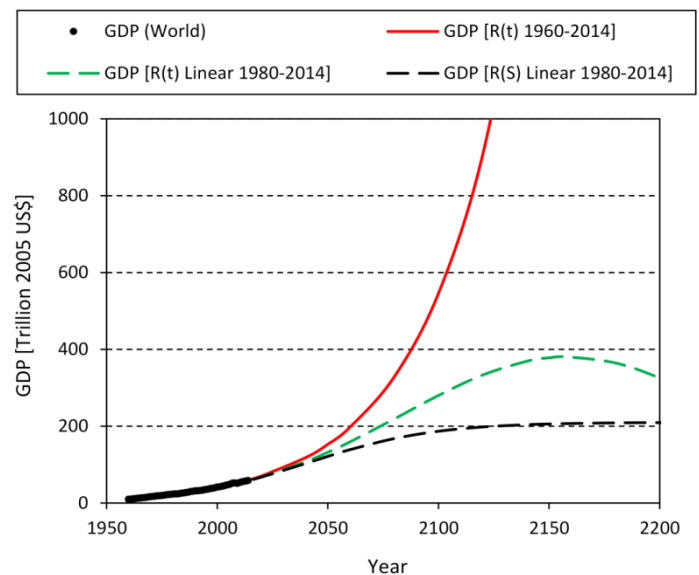

Figure 9. The most likely trajectory, *GDP* [*R*(*t*) 1960-2014], obtained by fitting empirical growth rate between 1960 and 2014 and described by the eqn (16) is compared with two projections based on linear approximations of the empirical growth rate between 1980 and 2014 and described by the eqns (27) and (28).

Table 1. Summary of the three projections of growth of the world GDP

| Year | T1 Trillion 2005 US$ | σ | Δσ/Δt [%] | T2 Trillion 2005 US$ | σ | Δσ/Δt [%] | T3 Trillion 2005 US$ | σ | Δσ/Δt [%] |
|---|---|---|---|---|---|---|---|---|---|
| 2000 | 41 | 1 | | 41 | 1 | | 41 | 1 | |
| 2014 | 58 | 1.4 | 3.7 | 58 | 1.4 | 3.7 | 58 | 1.4 | 3.7 |
| 2050 | 153 | 3.7 | 9.5 | 132 | 3.2 | 6.3 | 125 | 3.1 | 4.6 |
| 2100 | 546 | 13.4 | 33.6 | 280 | 6.9 | 7.2 | 192 | 4.7 | 1.9 |
| 2150 | 1,942 | 47.6 | 119.4 | 378 | 9.3 | 1.5 | 211 | 5.2 | 0.4 |
| 2158 | 2,379 | 58.4 | 146.3 | 380 | 9.3 | 0.1 | 212 | 5.2 | 0.3 |
| 2200 | 6,909 | 169.5 | 424.7 | 325 | 8.0 | -5.9 | 215 | 5.3 | 0.1 |
| 2250 | 24,578 | 602.8 | 1510.8 | 178 | 4.4 | -7.2 | 216 | 5.3 | 0.0 |
| 2300 | 87,435 | 2144.6 | 5374.6 | 62 | 1.5 | -3.9 | 216 | 5.1 | 0.0 |

T1, T2 and T3 – Calculated trajectories of the word GDP expressed in trillions of 2005 US$.

T1, labelled as GDP [R(t) 1960-2014] in Figure 9, was calculated using the growth rate given by the eqn (4) reproducing the growth rate data between 1960 and 2014. The trajectory is described by the eqn (16).

T2, labelled as GDP [R(t) Linear 1980-2014] in Figure 9, was calculated using *linear, time-dependent*, growth rate reproducing the empirical growth rate between 1980 and 2014. The trajectory is described by the eqn (27).

T3, labelled as GDP [R(S) Linear 1980-2014] in Figure 9, was calculated using *linear, GDP-dependent*, growth rate reproducing the empirical growth rate between 1980 and 2014. The trajectory is described by the eqn (28).

σ – Economic stress factor defined by the eqn (29)

Δσ/Δt – Annual increase of the economic stress factor.

Table 2. Formulae for the growth-rate-generated distributions describing the projected growth of the world GDP [represented as $S(t)$] and their corresponding, empirically-determined, parameters

| T1 | T2 | T2 |
|---|---|---|
| $S(t) = C \exp\left[\dfrac{t}{a} + \dfrac{1}{ra}\ln(a - be^{-rt})\right]$ | $S(t) = C \exp\left[at + 0.5bt^2\right]$ | $S(t) = \left[Ce^{-at} - \dfrac{b}{a}\right]^{-1}$ |
| $C = 5.650 \times 10^{-22}$ | $C = 1.095 \times 10^{-180}$ | $C = 1.093 \times 10^{29}$ |
| $a = 3.940 \times 10^{1}$, | $a = 3.895 \times 10^{-1}$ | $a = 3.539 \times 10^{-2}$ |
| $b = 3.787 \times 10^{42}$ | $b = -1.805 \times 10^{-4}$ | $b = -1.641 \times 10^{-4}$ |
| $r = 4.836 \times 10^{-2}$ | | |

$S(t)$ – GDP. See also the note under the Table 1.

It would be incorrect to describe the logistic limit to growth as the carrying capacity because the carrying capacity is not determined by fitting GDP data but by the ecological conditions. The predicted limit is just a limit defined by this particular linear approximation of the growth rate shown in Figure 8.

If the ecological limit for the global economic growth is below the predicted trajectory, economic growth will not be ecologically supported. The same applies to the predicted maximum based on the linear approximation of the growth. The predicted maximum is not necessarily the ecologically-sustainable maximum. If the ecological limit for the economic growth is below the predicted maximum or below the predicted logistic level, the growth rate should be now decreasing faster than indicated by the straight lines in Figure 8.

However, economic growth might become unstable and unmanageable even if the annual GDP value is below the ecological limit. In determining the sustainability of economic growth one has to consider not only the ecological limit but also the *economic stress factor*, $\sigma$, which can be defined as the ratio of two GDP values, at a time $t$ and at a selected time $t_0$:

$$\sigma \equiv \frac{GDP(t)}{GDP(t_0)}. \tag{29}$$

This factor describes approximately how much effort is required to generate a given level of the GDP when compared with a reference GDP. Thus, for instance, the world GDP increased from $41 trillion (2005 US$) in 2000 to $58 trillion in 2014. Even though the growth rate during that time decreased from 2.8% to 2.7%, the world GDP increased by 40%, reflecting the necessary larger effort to propel the economic growth.

If the world economic growth continues along the most likely trajectory (T1), the economic stress factor will increase from 1.4 in 2014 to 3.7 in 2050 and to 13.4 in 2100. Our effort to maintain the economic growth in 2014 was only 40% higher than in 2000 but by 2050 the intensity of economic activities would have to quadruple and by 2100 it would have to be 13 times the intensity of the year 2000. It might be impossible to cope with such a large stress and the economic growth might become unmanageable even during the current century, unless it is diverted to a slower trajectory.

In contrast, if the growth rate could decrease linearly with *time* along the most likely trajectory presented in the top section of Figure 8, then the expected economic stress factor could increase to 3.2 in 2050 and to 6.9 in 2100, which is still high but significantly lower than 13.4 for the T1 trajectory. However, if the growth rate could decrease linearly with the *size* of the GDP, as shown in the lower section of Figure 8, the economic stress factor could increase to a maximum of only about 5.3. This level of stress might be manageable.

For the T3 trajectory, when the growth approaches its limit, the growth rate continues to decrease with the size of the GDP but slowly. If, at any stage, the growth rate became constant, the growth of the GDP would be converted to exponential. However, if the growth rate started to increase, the growth of the GDP would also start to increase along a non-exponential trajectory. If the growth rate started to increase linearly with the size of the GDP, the growth of the GDP would start to follow a pseudo-hyperbolic trajectory (Nielsen, 2015b) escaping to infinity at a fixed time. Consequently, even for this safest trajectory, there might be a fine balance between the economic security and the economic crisis.

Equally important is to consider the annual increase in the economic stress factor. The stress factor might be relatively high but if it increases slowly with time such an increase might be tolerated. On the other hand, if the economic stress factor increases too fast, economic growth might become unstable.

The annual increase of the economic stress factor was 3.7% in 2014 but for the most likely trajectory (T1) it would increase to 9.5% in 2050 and to 33.6% in 2100, indicating that it might be hard or even impossible to follow this trajectory even during the current century. The projection is more promising for the T2 trajectory (corresponding to the growth rate decreasing linearly with *time*) showing a 6.3% annual increase in 2050 and only 7.2% in 2100. These figures are not too far from the 3.7% value in 2014 and consequently this growth might be still manageable. The best outcome is for the T3 trajectory (corresponding to the growth rate decreasing linearly with the *size* of the GDP) for which the maximum annual increase of the economic stress factor is 4.6% in 2050 and decreases to 1.9% in 2100, which is significantly lower than the corresponding value of 3.7% in 2014.

If we could control global economic growth, this is the trajectory we should follow. However, because we cannot control it, the most likely outcome is not promising. The current, approximately exponential, growth will have to be terminated but its spontaneous termination is undesirable because it would be most likely triggered by a global economic crisis rather than by secure economic conditions.

**Summary and conclusions**

We have analysed the world economic growth using the data of the World Bank (2015) between 1960 and 2014. We have demonstrated that, whether expressed as the function of time or of the size of the GDP, the growth rate has followed a well-defined trajectory, which is now approaching its asymptotic limit of approximately 2.5% per year. Between 1980 and 2014, the growth rate decreased from approximately 3.2% to 2.7%. The current growth is merging with the exponential growth, which is undesirable because after a certain time such a growth becomes unsustainable.

In 2014, the world GDP was $58 trillion 2005 US$. Should the economic growth continue along the most likely trajectory determined by the growth rate between 1960 and 2014, the world GDP would increase over 9-folds around the year 2100 (when compared with the GDP value in 2014), 33-folds around 2150, 120-folds around 2200, 4,237-folds around 2250 and 15,075-folds in 2300. These long term projections illustrate that such a growth cannot be tolerated over a long time. The current growth trajectory will have to be terminated, most likely spontanously but the question is whether such a spontaneous termination will be safe.

The growth of the world population is now slowing down but it is still rapidly increasing (Nielsen, 2015c). Human individual demands also continue to increase. It is possible that the growth of the world population will reach a maximum but a maximum in the individual demands appears to be far in the distant future. The two trends keep propelling the current economic growth and there is no change to favourable conditions in sight.

It is even likely that the economic growth rate will stop hovering around its asymptotic value and will start to increase to cope with the still-increasing demands. Such an outcome could be catastrophic because it would quickly lead to a rapid economic growth and even to singularity (Nielsen, 2015b). A trajectory characterised by singularity is potentially even more dangerous than the exponential growth because at a certain fixed time it escapes to infinity. The threat of the global economic crisis cannot be neglected.

For a safer economic growth, the growth rate should be now *decreasing below its asymptotic trajectory*. For the growth rate *decreasing linearly with time*, the economic growth would reach a maximum at a certain time and would start to decrease. For the growth rate *decreasing linearly with the size* of the GDP, the economic growth would approach asymptotically a certain limit. Examples of calculations based on a modest assumption that the growth rate would be decreasing along linear trajectories determined by fitting the empirical growth rate between 1980 and 2014 show that the world GDP would reach a maximum of $380 trillion (of 2005 US$) in 2158 or would be close to its asymptotic limit of $216 trillion around 2200. The world economic growth has now reached a critical stage and its future is not reassuring.